# A Detailed Investigation of the Proposed NN Serpentis Planetary System

J. Horner[1], R. A. Wittenmyer[1], T. C. Hinse[2,3] & C. G. Tinney[1]

[1] Department of Astrophysics and Optics, School of Physics, University of New South Wales, Sydney 2052, Australia

[2] Korea Astronomy and Space Science Institute, 776 Daedeokdae-ro, Yuseong-gu, 305-348, Daejeon, Republic of Korea (South)

[3] Armagh Observatory, College Hill, BT61 9DG, NI, UK

**Abstract**
The post-main sequence eclipsing binary NN Serpentis was recently announced as the potential host of at least two massive planetary companions. In that work, the authors put forward two potential architectures that fit the observations of the eclipsing binary with almost identical precision. In this work, we present the results of a dynamical investigation of the orbital stability of both proposed system architectures, finding that they are only stable for scenarios in which the planets are locked in mutual mean motion resonance. In the discovery work, the authors artificially fixed the orbital eccentricity of the more massive planet, NN Ser(AB) c, at 0. Here, we reanalyse the observational data on NN Serpentis without this artificial constraint, and derive a new orbital solution for the two proposed planets. We detail the results of further dynamical simulations investigating the stability of our new orbital solution, and find that allowing a small non-zero eccentricity for the outer planet renders the system unstable. We conclude that, although the original orbits proposed for the NN Serpentis planetary system prove dynamically feasible, further observations of the system are vital in order to better constrain the system's true architecture.

**Introduction**
In recent years, the search for planets orbiting other stars has yielded a vast number of planetary bodies moving on a wide range of orbits around a hugely diverse variety of stars (e.g. Mayor et al. 1995; Johnson et al. 2010; Tinney et al., 2011; Doyle et al. 2011; Anglada-Escudé et al. 2012, Muirhead et al., 2012). In many ways, the most unusual recent discoveries have featured the detection of planets in elderly and evolved stellar systems. Whilst some such planets have been detected orbiting solitary evolved stars (e.g. Hatzes et al., 2005; Sato et al. 2010; Johnson et al. 2011), the most startling have been those planets detected through variations in the timing of eclipses between very tightly bound components of evolved binary star systems (e.g. HW Virginis, Lee et al., 2009; HU Aquarii, Qian et al., 2010; and NN Serpentis, Beuermann et al., 2010).

It has recently become evident, however, that the story might not be as clear-cut as previously thought. In at least two cases (HU Aqr and HW Vir), dynamical analyses have shown that proposed planetary systems do not stand up to scrutiny (Horner et al. 2011, 2012b). Horner et al. (2011) were the first to bring this matter to light, performing detailed dynamical simulations that allowed them to map the stability of the planets proposed in the HU Aqr system as a function of their orbital elements. They found that it was highly improbable that the observed signal was the result of perturbations due to planetary mass companions. The orbits proposed for those planets proved dynamically unstable on such short timescales that the likelihood of observing such a system before it disintegrated is vanishingly small. Re-analysis of the observations of HU Aqr (Wittenmyer et al., 2012a**,** Horner et al., 2012a**,** Hinse et al., 2012a) have shown that, whilst HU Aqr clearly displays variations in the timing of its eclipses, those variations must be the result of something other than the influence of two planetary companions. Gozdziewski et al. (2012) have recently presented new observations of the HU Aqr system, and performed an extensive analysis of the newly enlarged observational dataset. In that work, they find that a wide range of 2-planet configurations are capable of fitting the observational data, but that the best fit is obtained when the presence of just a single planet is invoked. Recently, Horner et al. (2012b) have performed a similarly detailed analysis of the proposed planets orbiting the evolved binary HW Vir. Once again, the proposed planets do not stand up to dynamical scrutiny. Indeed, in the case of HW Vir, the system proved dynamically unstable on timescales of just a few hundred years for all allowed orbital solutions within three sigma of the nominal best-fit orbit.

It is becoming ever more apparent that, whenever multiple planets are suspected to orbit a given star, their long term dynamical behaviour must be considered prior to their presence being stated with any certainty. Indeed, such dynamical studies are now becoming a key component of exoplanet discovery papers (e.g. Robertson et al., 2012(a,b), Anglada-Escudé et al. 2012, Wittenmyer et al., 2012b, Hinse et al., 2012b). It is therefore timely to perform a study of the dynamics of the proposed NN Serpentis planetary system, in order to confirm that the planets therein are indeed dynamically feasible.

In Section 2, we review the NN Serpentis planetary system, as proposed in Beuermann et al., 2010, before describing our dynamical investigation of the stability of that system in Section 3. In Section 4, we describe a new analysis of the observational data on NN Serpentis, before examining the stability of our new and improved orbital solution for the system in Section 5. Finally, we present our conclusions in Section 6.

**The NN Serpentis System**
Haefner et al. (2004) present an overview of the physical parameters of the NN Serpentis system – they find that orbital period of the stars is 3h 7m, the mass of the primary is 0.54±0.05 $M_\odot$ and the mass of the secondary 0.150±0.008 $M_\odot$. It is interesting to note that the primary's relatively high temperature (57000±3000 K) suggests that its evolution to become a white dwarf occurred relatively recently (i.e. only approximately $10^6$ years ago; Wood 1995). Unlike HU Aqr, in which the eclipses are muddied by the influence of accretion between the primary and secondary stars, NN Serpentis is considered to be a pre-cataclysmic variable. At the current epoch, although the secondary is slightly ellipsoidal, matter is not being transferred between the secondary and primary. The secondary does, however, display a significant temperature gradient as a result of heating by the white dwarf primary, with the temperature of the side tidally locked to face the primary heated by ~4200K over that of the side facing away from the white dwarf. This leads to an ellipsoidal variation in the out-of-eclipse brightness of the system, as the phase of the secondary star varies during its orbit. For more details of the NN Serpentis binary system, we refer the interested reader to Haefner et al., 2004, and references therein.

In 2010, Beuermann et al. reported that long-term, systematic variations in the timings of NN Serpentis' eclipses could be explained as being the result of perturbations from two massive planets orbiting in the system. They put forward two models that offered almost equally good fits to the observed data. The first featured planets close to, or trapped within, mutual 2:1 orbital mean-motion resonance, with periods of ~15.5 (planet c) and ~7.7 years (planet d). Their second solution involved planets close to, or within, mutual 5:2 resonance, with periods of ~16.7 and 6.7 years (planets c and d, respectively). The key parameters of their two best models are presented in Table 1, which is based on their Table 2. It is important to note that, as part of their orbital fitting procedure, they held the orbital eccentricity of the more massive planet, planet c, fixed at zero, and treated the other orbital parameters as free parameters. This resulted in two possible model architectures for the system that were essentially equally good, as detailed in Table 1. Both solutions invoked a moderate eccentricity for the orbit of planet d, the less massive and shorter period of the two proposed in that work.

| Solution | $a_c$ (AU) | $a_d$ (AU) | $e_c$ | $e_d$ | $m_c \sin i$ ($M_{Jup}$) | $m_d \sin i$ ($M_{Jup}$) |
|---|---|---|---|---|---|---|
| 2:1 | 5.38±0.20 | 3.39±0.10 | 0.00 | 0.20±0.02 | 6.91±0.54 | 2.28±0.38 |
| 5:2 | 5.66±0.06 | 3.07±0.13 | 0.00 | 0.23±0.04 | 5.92±0.40 | 1.60±0.27 |

**Table 1:** The two solutions obtained by Beuermann et al., 2010. For more details on the solutions themselves, we refer the reader to Table 2 of that work. Here, *a* is the semi-major axis of the planet in question, and *e* its orbital eccentricity. Note that the eccentricity of planet c was held fixed at 0.00 in both cases, while the other parameters were allowed to vary as free parameters. The mass of the two planets, for a given solution, is given by *m* sin *i*, in units of Jupiter's mass. The orbits of the planets in the first solution (labelled 2:1) lie close to, or within, mutual 2:1 mean-motion resonance, while those in the second solution (labelled 5:2) lie close to mutual 5:2 mean-motion resonance.

In the discovery work, the authors mention that "*the probable detection of resonant motion with a period ratio of either 2:1 or 5:2 is a major bonus, which adds to the credence of the two-planet model.*" Although they briefly mention a short *N*-body simulation of their best-fit 2:1 resonant solution, it is not clear whether they performed any significant long-term integration to investigate the stability of their solutions. It is clear from studies of our own Solar system that, just because two objects are currently exhibiting resonant behaviour, such behaviour does not necessarily result in those objects being dynamically stable on astronomically long timescales (e.g. Horner & Lykawka, 2010; Horner et al., 2012c; Horner, Müller & Lykawka, 2012; Dvorak et al., 2012). As such, it is clearly important to consider the long-term evolution of the system, rather than simply checking whether it is currently exhibiting resonant behaviour.

**The Dynamics of the Proposed Planets**
In light of other circumbinary planets (HW Vir and HU Aqr) that have since been shown to be dynamically unfeasible, there is sufficient motivation to examine the dynamical stability of the proposed planets in the NN Serpentis system. Beyond this motivation, the fact that Beuermann et al. (2010) present two distinctly different potential orbital architectures for the NN Serpentis system yields an additional reason to study the dynamics of this system. Can we use a detailed dynamical analysis to help determine which of the proposed architectures is more reasonable for the planetary system?

To study the dynamics of the proposed NN Serpentis planetary system, we followed a now well-established route. Building on the work of Marshall et al. (2010) and Horner et al. (2011), we used the *Hybrid* integrator within the *n*-body dynamics package MERCURY (Chambers, 1999) to perform two suites of detailed dynamical simulations of the NN Serpentis system, one for each of the orbital architectures proposed by Beuermann et al. (2010). As in our earlier work, we held the initial orbit of the planet with the most tightly constrained orbital elements (NN Serpentis (AB) c – hereafter simply "planet c") at the nominal best-fit solution, and varied the initial orbit of the other planet (NN Serpentis (AB) d – hereafter simply "planet d"), such that we tested orbits spanning the full 3 sigma error range in that planet's semi-major axis, *a*, eccentricity, *e*, longitude of periastron, *ω*, and mean anomaly, *M*. The masses of the two planets were set at the nominal *m sin i* values given in Beuermann et al. (2010) for each of the scenarios tested. Since the gravitational influence of the two planets is a function of their mass, then one would expect that the greatest likelihood of finding stable solutions for the system would occur at the lowest planetary masses (ignoring the potential influence of secular resonances, which can be the cause of otherwise unexpected instability e.g. Horner & Jones, 2008).

For each of the scenarios proposed by Beuermann et al., (2010), a total of 91,125 possible orbits were tested for planet d, forming a 4-dimensional cube in *a-e-ω-M* space. 45 unique and evenly spaced values of *a* and *e* were tested. At each of these 2025 *a-e* pairs, we considered 15 unique values of *ω*, and 3 values of *M*, all spread evenly across the 3-sigma uncertainties in those elements. The simulations were run for a period of 100 million years, with the mass of the two central stars combined to one object located at the system barycentre (as is standard in such work), and the orbital evolution of the two planets was followed until they collided with one another, were ejected from the system, or collided with the central body. As in our earlier works, the ejection distance was set to 10 AU, since for either planet to reach that barycentric distance, significant mutual interactions must occur between the two planets.

*Original Solution I: Planets close to 2:1 mean-motion resonance*
The first orbital architecture proposed by Beuermann et al. (2010) placed the planets in the NN Ser system very close to their mutual 2:1 mean-motion resonance, and is given in Table 1. The 2:1 orbital solution placed planet c on an orbit with *a* = 5.38 AU, and a mass 6.91 times that of Jupiter. The orbital eccentricity of that planet was held fixed in their calculations, which assumed its orbit to be circular. This resulted in a best-fit orbit for planet d with *a* = 3.39 AU, and *e* = 0.20, and a mass for that planet of 2.28 $M_J$. 45 unique semi-major axes were tested for planet d, ranging from 3.09 to 3.69 AU in equal sized steps. At each of these initial semi-major axes, 45 unique eccentricities were considered, ranging from 0.14 to 0.26, again in even steps. Fifteen initial longitudes of periastron were tested that ranged between 62 and 86 degrees, around a nominal best fit value of 74 degrees, and three widely-spaced initial mean anomalies were tested, namely 330 degrees, 0 degrees, and 30 degrees.

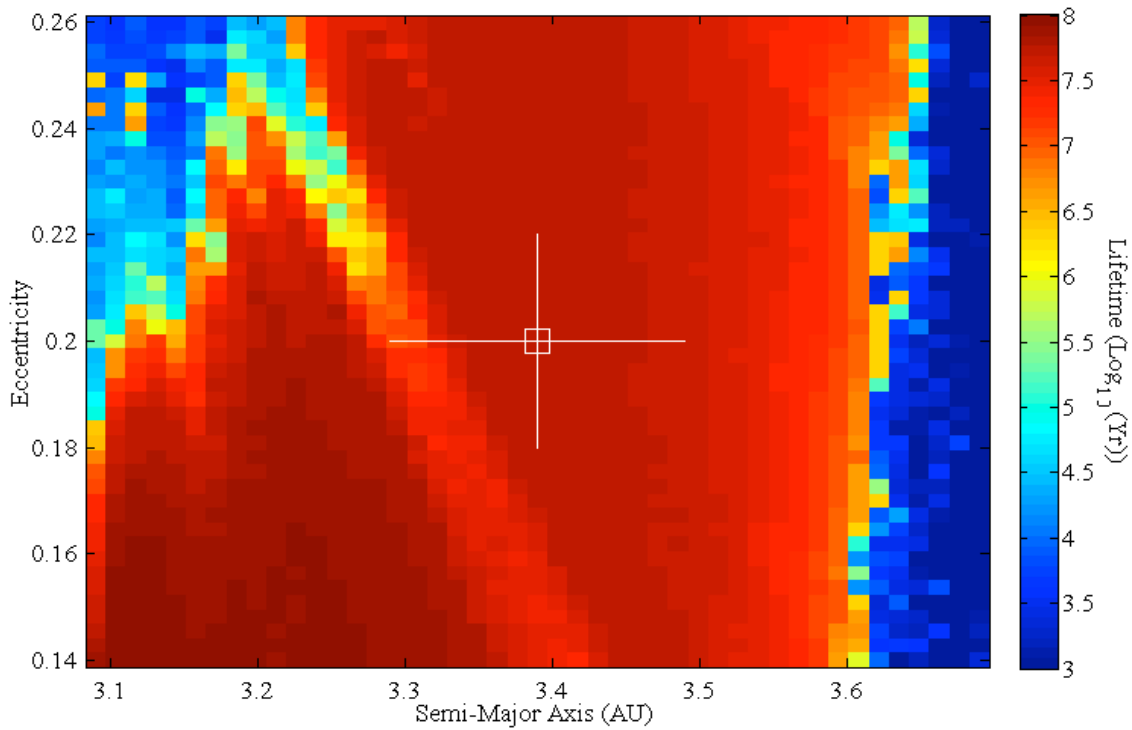

**Figure 1:** The dynamical stability of the NN Serpentis system for the Beuermann et al. (2010) scenario in which the planets are in mutual 2:1 resonance, as a function of the initial semi-major axis and eccentricity of the orbit of planet d. The nominal best-fit orbit for that planet is marked by the open square, while the 1-sigma errors on that value are shown by the lines radiating from that point.

In Figure 1, we present the results of our dynamical simulations of this proposed architecture for the NN Serpentis planetary system. As has been seen in previous studies of potentially resonant exoplanets (e.g. Robertson et al., 2012b), resonant interactions can significantly affect the stability of the planetary system in question. Here, the broad 2:1 mean motion resonance between the two planets offers a large region of stability, which is bounded by a sharp transition to highly unstable solutions. In stark contrast to the proposed planetary systems around HW Vir and HU Aqr, we find that the proposed planetary system around NN Serpentis is dynamically feasible, at least for the scenario featuring planets trapped in mutual 2:1 mean-motion resonance. The only regions of strong instability lie well away from the nominal best-fit orbit for planet d, and highly stable orbital solutions can be found throughout the central ±1 sigma region. We found that the initial longitude of periastron for planet d (as shown in figure 2) had relatively little effect on the stability (or instability) of the solutions tested – aside from the region between $a \sim 3.50 - 3.65$ AU, at the outer edge of the region of stability offered by the 2:1 MMR. In that region, orbits with smaller initial $\omega$ tended to be significantly more stable than those at larger $\omega$, with a sharp divide between stable and unstable orbits progressing in a roughly linear manner from the maximum value of $\omega$ (86 degrees) at ~3.5 AU to the minimum $\omega$ tested (62 degrees) at ~3.65 AU.

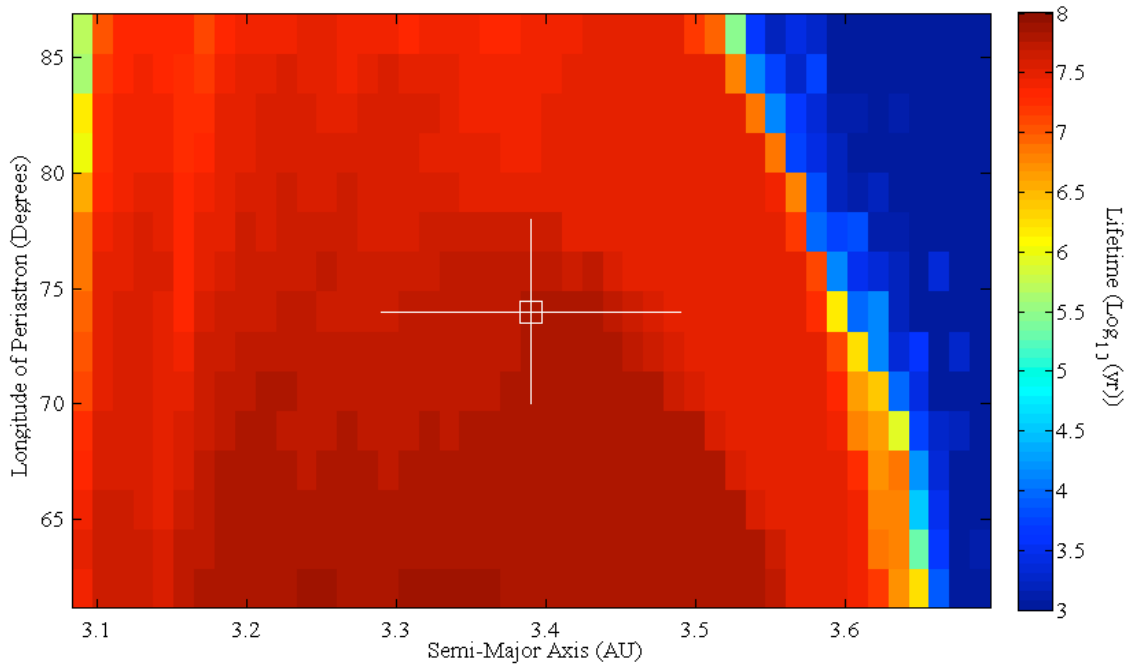

**Figure 2:** The dynamical stability of the NN Serpentis system for the 2:1 resonant scenario, as a function of the initial semi-major axis and longitude of periastron, $\omega$, of the orbit of planet d. The nominal best-fit orbit for that planet is marked by the open square, while the 1-sigma errors on that value are shown by the lines radiating from that point. Highly stable solutions are found throughout the central ±1-sigma region, with very little variation in mean lifetime as a function of $\omega$. At larger semi-major axes, $\omega$ begins to play an important role in determining the stability of the orbit of planet d, with smaller values of $\omega$ offering significantly greater stability than higher values.

In total, we tested three unique initial values of mean anomaly for planet d. Planet c was placed on an orbit at an initial mean anomaly of 213 degrees, and each $a$-$e$-$\omega$ solution we considered for planet d was then tested with initial mean anomalies of 330, 0 and 30 degrees (an initial setup based on the statement in Beuermann et al. (2010) that "*Periastron passage of NN Ser (ab)d occurred last on JD′ ≃ 4515. At that time NN Ser (ab)c was at longitude 213°*"). Our estimated error for the mean anomaly of ±10 degrees is doubtless a very conservative estimate, but does result in an interesting observation – whilst the allowed orbits for planet d are stable across the whole ±1 sigma range of semi-major axes and inclinations for initial mean anomalies of 330 and 0 degrees (with planet d initially at M = 213 degrees), placing planet d at an initial mean anomaly of 30 degrees leads to a drastic destabilisation of orbits within the region around the best-fit solution. This can be clearly seen in figure 3, below. It is clear that, if the initial mean anomaly of planet d's orbit is significantly different to that suggested in Beuermann et al. (2010), then the 2:1 commensurability between the orbital periods of the two planets can lead to highly destructive dynamical evolution. Fortunately, the best-fit solution for the two planets lies at mean anomalies that favour stable, rather than unstable, solutions, something that would be expected if the observed variation in eclipse timings in NN Serpentis were the result of perturbations by two planets trapped in mutual 2:1 resonance.

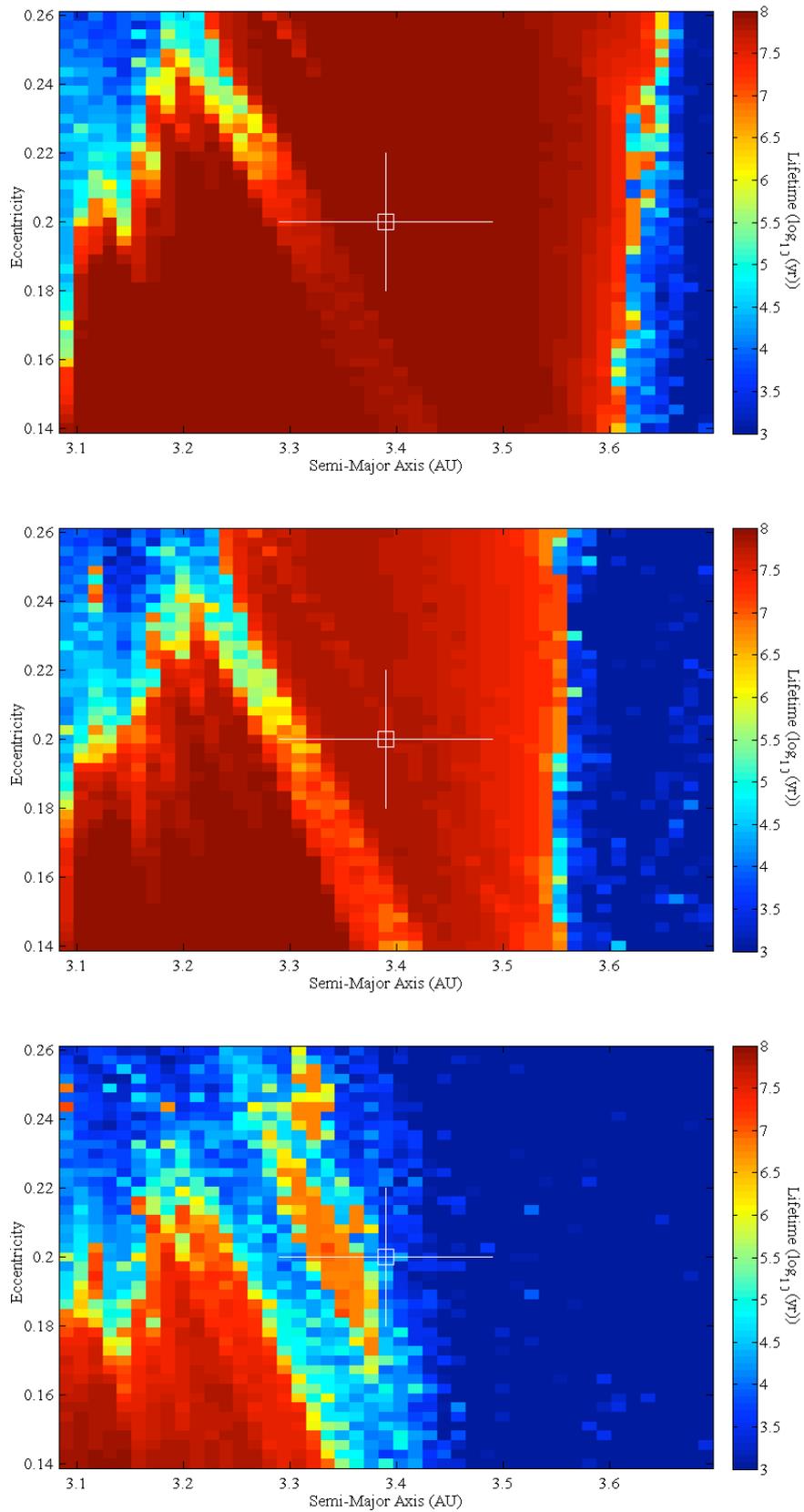

**Figure 3:** The dynamical stability of the 2:1 resonant solution for the NN Serpentis planetary system, as proposed by Beuermann et al. (2010), as a function of semi-major axis and eccentricity, for three distinct values of initial mean anomaly for planet d. The upper panel shows the scenario in which the initial mean anomaly of planet d is set to 330 degrees, with that of planet c set to 213 degrees. The central panel has planet d at an initial $M = 0$ degrees, while the lower panel shows the situation where *M* = 30 degrees. It is clear that, if the initial mean anomaly for planet d is significantly different from the best fit value, it is possible to induce extreme instability in the planetary system, as the 2:1 commensurability between the orbits of the two planets shifts from a protective, stable scenario to a destructive, unstable scenario.

### *Original Solution II: Planets close to 5:2 mean-motion resonance*

The second orbital architecture proposed by Beuermann et al. (2010) placed NN Serpentis' planets very close to their mutual 5:2 mean-motion resonance. That solution placed planet c on an orbit with *a* = 5.66 AU, with a mass of 5.92 $M_J$. Once again, the orbital eccentricity of that planet was held fixed at zero in their calculations. This resulted in a best-fit orbit for planet d with *a* = 3.07 AU, and *e* = 0.23, and a mass for that planet of 1.60 $M_J$. 45 unique semi-major axes were tested for planet d, ranging from 2.68 to 3.46 AU in equal sized steps. At each of these initial semi-major axes, 45 unique eccentricities were considered, ranging from 0.11 to 0.35, again in even steps. Fifteen initial longitudes of periastron were tested that ranged between 52 and 94 degrees, around a nominal best fit value of 73 degrees, and three widely-spaced initial mean anomalies were tested, namely 330 degrees, 0 degrees, and 30 degrees.

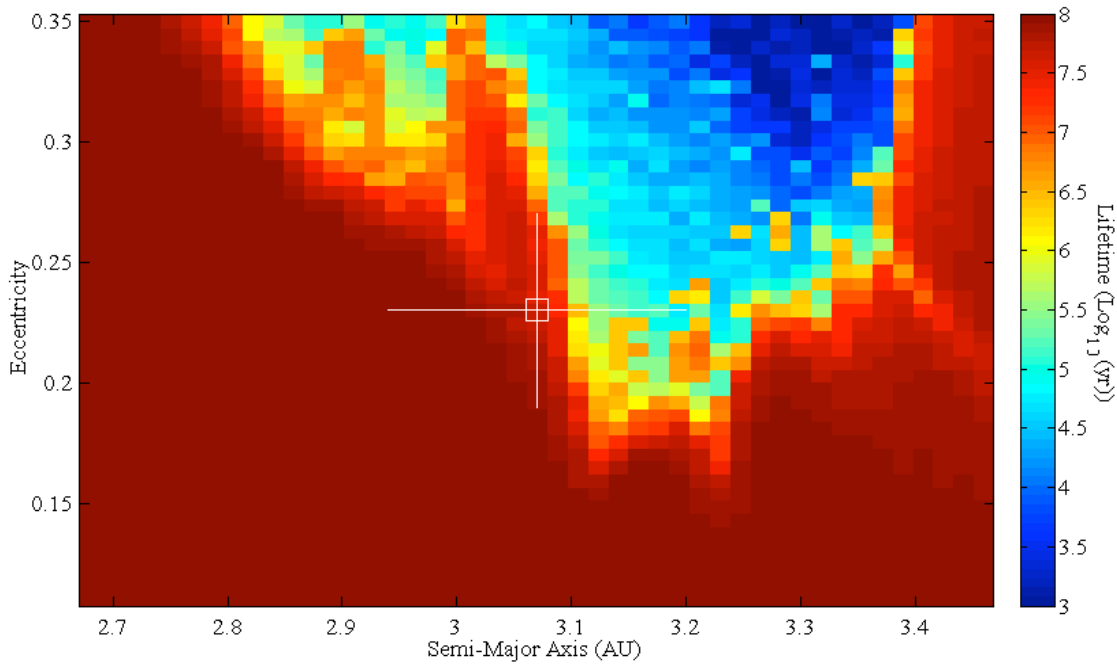

**Figure 4:** The dynamical stability of the NN Serpentis system for the Beuermann et al. (2010) scenario in which the planets are in mutual 5:2 resonance, as a function of the initial semi-major axis and eccentricity of the orbit of planet d. Once again, the nominal best-fit orbit planet d is marked by the open square, while the 1-sigma errors on that value are shown by the lines radiating from that point.

In Figure 4, we present the results of our simulations investigating the stability of their second scenario. This scenario featured planets of slightly lower mass and wider separation than was the case for their 2:1 resonant architecture. Taken together, this might infer that the scenario would offer better prospects for stability. However, as is apparent from Figure 4, this is not necessarily the case. Indeed, it turns out that the nominal best-fit orbit for planet d places it right on the boundary between dynamically stable orbits and those that are heavily destabilised by the influence of the 5:2 mean-motion resonance. As the initial eccentricity of the orbit of planet d is increased, the 5:2 mean-motion resonance becomes markedly more disruptive, and broadens noticeably. As a result, only about half of the orbits within the inner ±1 sigma region of allowed orbital element space prove to be dynamically stable. Whilst this certainly does not rule out this proposed orbital architecture for the NN Serpentis planets, it does suggest that, in this scenario, the orbit of planet d is likely located at a somewhat smaller orbital radius than is indicated on the basis of the observational results alone, and/or at a somewhat smaller orbital eccentricity.

In contrast to the case of the 2:1 resonant scenario discussed above, we found that the stability (or otherwise) of the orbits tested for planet d in this case were entirely unaffected by variations in longitude of perihelion and mean anomaly – in this case, regions of stability and instability were solely functions of the initial semi-major axis and eccentricity tested.

**Re-analysing the Observational Data**
In light of the fact that Beuermann et al. (2010) artificially set the orbital eccentricity of planet c to be zero, it is interesting to consider the effect that this might have had on the final solutions they obtain. We have reanalysed the discovery data, removing this constraint, to see whether this resulted in a significantly different orbital solution. Repeating the analysis method used in Wittenmyer et al. (2012a), we performed Keplerian orbit fits to the timing data given in Beuermann et al. (2010). We applied their linear ephemeris to obtain the (O-C) residuals, which we then modelled as two Keplerian orbits. To explore a large parameter space and ensure a globally optimised solution, we used a genetic algorithm (Charbonneau et al. 1995) as in our previous work (Wittenmyer et al. 2012a, Wittenmyer et al. 2011, Tinney et al. 2011). The fitting process is analogous to the well-established techniques used to fit orbits to radial-velocity data – the chief difference is that the velocity semi-amplitude $K$ (with units of metres per second) is now a timing semi-amplitude measured in seconds. We ran the genetic algorithm for 100,000 iterations, each consisting of ~1000 individual trial fits. The best-fit set of parameters is thus the result of ~$10^8$ trial fits. The parameters of the best 2-planet solution obtained by the genetic algorithm were then used as initial inputs for the GaussFit code (Jefferys et al., 1987), a nonlinear least-squares fitting routine, to obtain a final two-Keplerian model fit. The results are given in Table 2.

| Parameter | Inner Planet (d) | Outer Planet (c) |
|---|---|---|
| **Orbital Period (days)** | 2605±124 | 5571±67 |
| **Amplitude (sec)** | 9.5±1.4 | 26.3±0.6 |
| **Eccentricity** | 0.05±0.02 | 0.22±0.06 |
| **ω (°)** | 152±50 | 39±15 |
| **$T_0$ (JD-2400000)** | 58029±173 | 53155±217 |
| **a (AU)** | 3.20±0.42 | 5.32±0.28 |
| **m sin i ($M_{Jup}$)** | 4.00±0.62 | 6.71±0.41 |
| **Reduced $\chi^2$** | 0.75 | |
| **RMS (sec)** | 3.23 | |

**Table 2:** The results of our best-fit two-Keplerian model fit for the planetary system around NN Serpentis.

**The Dynamics of our New Solution**
As before, we performed a suite of 91,125 integrations, sampling the full three sigma range of plausible orbits for planet d. As in our earlier integrations, we held the initial orbit of planet c fixed at its nominal best-fit location ($a$ = 5.3167 AU, $e$ = 0.215). The best-fit solution for the orbit of planet d placed the planet at $a$ = 3.20 AU, $e$ = 0.05. In our integrations, we once again used the derived *m sin i* values for the masses of the two planets, with planet c having a mass of 6.705 $M_J$.

planet d had a mass of 4.003 $M_J$ – significantly greater than that obtained in either of the solutions presented by Beuermann et al., 2010. Once again, 45 unique semi-major axes were tested for planet d, ranging from 1.94 to 4.46 AU in equal sized steps. At each of these initial semi-major axes, 45 unique eccentricities were considered, ranging from 0.000 to 0.103, again in even steps. Fifteen initial longitudes of periastron were tested that ranged between 2.4 and 301.6 degrees, around a nominal best fit value of 152 degrees, and three widely-spaced initial mean anomalies were tested, namely 334.7 degrees, 46.4 degrees, and 118.1 degrees.

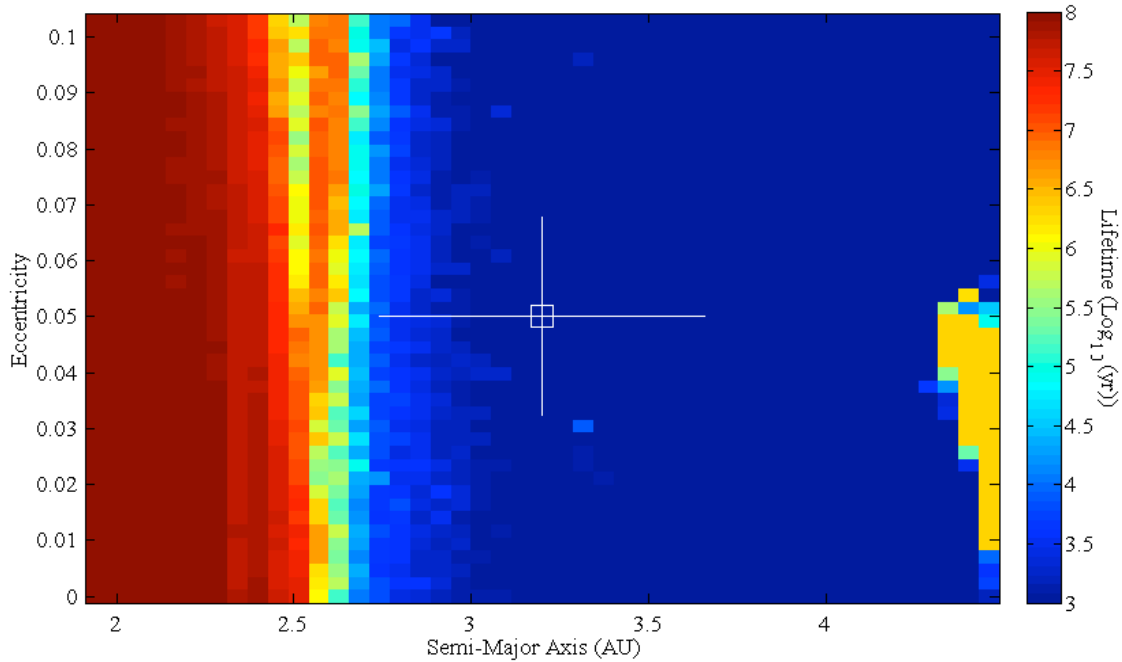

**Figure 5:** The dynamical stability of the NN Serpentis system as a function of the eccentricity and semi-major axis of planet d, for the new orbital solution derived in this work. The nominal best-fit orbit for planet d is marked by the hollow box, while the one sigma errors on that value in *a* and *e* are shown by the crosshairs. As a result of the dynamically excited orbit derived for planet c (*e* ~ 0.22), the best-fit solution for planet d now falls in a broad region of extreme instability.

In Figure 5, we present the results of our simulations investigating the stability of the new orbital solution for the NN Serpentis planetary system derived in this work. As a result of the removal of the constraint that forced the orbit of planet c, the more massive planet, to be circular, our solution features an orbit for that planet that has an eccentricity of almost 0.22, resulting in a periastron distance of just 4.173 AU. At that distance, planet c has an instantaneous Hill radius of 0.6215 AU. In our earlier work, we found that orbital solutions for exoplanets that feature orbits approaching one another more closely than ~3 Hill Radii are typically very unstable (Horner et al., 2011). It is no surprise, therefore, that orbits for planet d with semi-major axes greater than ~2.3 AU start to display significant dynamical instability. As was the case for the 5:2 scenario discussed above, the stability or instability of the our solutions was found to be independent of the initial mean anomaly and longitude of periastron of planet d. Once again, the regions of stability and instability were solely functions of the initial semi-major axis and eccentricity tested.

Whilst our results do show a broad region of stability for the NN Serpentis system, that region is located at semi-major axes more than one sigma away from the nominal best-fit orbit. It is perhaps somewhat discouraging that simply allowing the orbital solution for planet c to become moderately eccentric is so disruptive to the stability of the planetary system. That said, we note that the formal uncertainties in Keplerian orbital parameters obtained from the covariance matrix of nonlinear least-squares fits tend to be underestimated. In addition, there is a bias against obtaining fitted eccentricities near zero, as demonstrated by O'Toole et al. (2009). Especially for data with sampling-induced gaps in orbital phase, fitting routines tend to inflate the eccentricity in order to minimise $\chi^2$. With these two findings in mind, it is wholly reasonable that the true eccentricity of planet c is smaller than the nominal best-fit e=0.22, which would result in the stable region evident in Figure 5 moving outwards to encompass the 1-sigma region. We also note that the difference in reduced $\chi^2$ between our revised orbital solution and those of Beuermann et al. (2010) is quite small; we obtain 0.75 compared to 0.78 (2:1 solution) and 0.80 (5:2 solution).

**Conclusions**

The presence of two planets orbiting the cataclysmic variable system NN Serpentis was proposed by Beuermann et al. (2010), based on observations that showed that the timing of the mutual eclipses between the component stars of that binary was varying in a periodic manner. That work put forward two scenarios by which the observed eclipse timing variations could be explained by the presence of two orbiting planets. In the first, the two proposed planets were close to the mutual 2:1 mean-motion resonance, whilst in the second, they were instead close to the mutual 5:2 mean-motion resonance.

In this work, we have performed a highly detailed dynamical analysis of the two orbital architectures proposed for the NN Serpentis system in the discovery work. We find that the proposed NN Serpentis planetary system is dynamically stable, for both scenarios, across a wide range of the allowed orbital element space. Indeed, in the case of the 2:1 resonant scenario, almost all orbital solutions within one sigma of the nominal best-fit case exhibit strong dynamical stability. In the case of the 5:2 resonant solution, the best-fit orbit lies on the boundary between highly stable and highly unstable regimes. Nevertheless, at least half of the solutions within one sigma of the best-fit orbit exhibit strong dynamical stability; with a still greater fraction being stable when one considers the full ±3 sigma range of plausible orbits.

In their analysis, Beuermann et al. (2010) artificially held the orbital eccentricity of the more massive of their planets, planet c, fixed at zero. In this work, we revisited the analysis of the observational data on NN Serpentis, removing this constraint. Once the eccentricity of planet c's orbit was able to vary freely, we found that the best fit to the observational data was obtained with a moderately eccentric orbit (e~0.22). On the basis of this result, we obtained a new orbital solution for the NN Serpentis planetary system. In our new solution, the eccentricity of the innermost planet, planet d, was found to be significantly lower than was proposed by Beuermann et al., whilst its mass was significantly higher.

We then performed a detailed dynamical analysis of the new NN Serpentis system described by our solution. As a result of the outermost planet's increased orbital eccentricity, the great majority of allowed orbits for planet d featured significant dynamical instability, as the two planets interacted strongly with one another as a result of planet c's reduced periastron distance. Orbital solutions exterior to ~2.5 AU for planet d were found to be highly unstable (including the entire ± one sigma region centred on the nominal best-fit orbit). This highlights the importance of orbital eccentricity in determining the dynamical stability of a planetary system.

Our results suggest that the proposed planetary system around NN Serpentis is dynamically feasible. As more observations are taken, it will be possible to obtain better constraints on the plausible orbits for the proposed planets, and so more conclusively distinguish between the three possible orbital solutions discussed here. Since it is well known that orbital fitting routines often have a tendency to select against solutions with very low orbital eccentricities (O'Toole et al. 2009), it is possible that allowing the orbital eccentricity of planet c to vary freely has caused the fitting process to gravitate to a more dynamically unfeasible solution (with a higher eccentricity).

**Acknowledgements**

The authors wish to thank the anonymous referee of this paper, who made a number of very helpful comments and suggestions. JH gratefully acknowledges the financial support of the Australian government through ARC Grant DP0774000. RW is supported by a UNSW Vice-Chancellor's Fellowship. TCH gratefully acknowledges financial support from the Korea Research Council for Fundamental Science and Technology (KRCF) through the Young Research Scientist Fellowship Program. The work was supported by iVEC through the use of advanced computing resources located at the Murdoch University, in Western Australia. Astronomical research at Armagh Observatory (UK) is funded by the Department of Culture, Arts and Leisure (DCAL), Northern Ireland, UK. TCH also acknowledges support from KASI registered under grant number 2012-1-410-02.